\documentclass[floatfix,aps,prd,eqsecnum,showpacs,amsmath,
nofootinbib,preprintnumbers]{revtex4}

\usepackage{graphicx}

\newcommand{\beq}{\begin{equation}}
\newcommand{\eeq}{\end{equation}}
\newcommand{\bea}{\begin{eqnarray}}
\newcommand{\eea}{\end{eqnarray}}

\def\laq{~\raise 0.4ex\hbox{$<$}\kern -0.8em\lower 0.62
ex\hbox{$\sim$}~}
\def\gaq{~\raise 0.4ex\hbox{$>$}\kern -0.7em\lower 0.62
ex\hbox{$\sim$}~}

\def \pa {\partial}
\def \ra {\rightarrow}

\def \b {\beta}
\def \a {\alpha}

\def \Ga {\Gamma}
\def \ga {\gamma}
\def \sg {\sigma}
\def \da {\delta}
\def \ep {\epsilon}
\def \r {\rho}

\def \Om {\Omega}

\def \fb {\overline \phi}
\def \ls {\lambda_{\rm s}}
\def \lp {\lambda_{\rm P}}
\def \na {\nabla}
\def \rg {\sqrt{-g}}
\def \rp {\sqrt{\ep (\nabla \phi)^2}}
\def \cl {{\cal L}_m}
\def \fpu {\dot{\phi}}
\def \fpp {\ddot{\phi}}
\def \d {{\rm d}}
\def \e {{\rm e}}

\begin{document}

\preprint{BA-TH/07-583}
\preprint{gr-qc/yymmnnn}

\title{Non-local dilaton coupling to dark matter: \\ 
cosmic acceleration and pressure backreaction}

\author{L. Amendola$^{1}$, M. Gasperini$^{2,3}$ and C. 
Ungarelli$^{4}$}

\affiliation{$^{1}$INAF/Osservatorio Astronomico di Roma,\\
Via Frascati 33, 00040 Monteporzio Catone (Roma), Italy\\
 $^{2}$Dipartimento di Fisica, Universit\`{a} di Bari, Via G. Amendola
173, 70126 Bari, Italy\\
 $^{3}$INFN, Sezione di Bari, Bari, Italy\\
 $^{4}$CNR -- Institute of Geosciences and Earth Resources,\\
Via G. Moruzzi, 1, 56124  Pisa, Italy}

\begin{abstract}
A model of non-local dilaton interactions, motivated by string duality symmetries, is applied to a scenario of ``coupled quintessence" in which the dilaton dark energy is non-locally coupled to the dark-matter sources. It is shown that the non-local effects tend to generate a backreaction which -- for strong enough coupling -- can automatically compensate the acceleration due to the negative pressure of the dilaton potential, thus asymptotically restoring the standard (dust-dominated) decelerated regime. This result is illustrated by analytical computations and numerical examples. 
\end{abstract}


\maketitle

\section {Introduction}
\label{sec1}

It is well known that the ``cosmological" duality symmetries of the low-energy string effective action -- for instance, scale-factor duality 
\cite{1,2} -- require  non-linear transformations mixing the metric and the dilaton field $\varphi$: as a consequence, those symmetries are generally lost in the presence of a non-trivial dilaton potential $V=V(\varphi)$ and/or non-trivial couplings of the dilaton to the matter sources. It is also known, however, that the duality properties of the string-cosmology equations are preserved in the presence of the potential provided $V$  depends on $\varphi$ through an appropriate ``shifted" variable $\fb$, defined by $\fb= \varphi- \ln \sqrt{-g}$, where $g$ is the determinant of the metric tensor. In that case the equations are duality-invariant even in the generalized case of global $O(d,d)$ transformations, both without \cite{3,4} and with \cite{5} matter sources. 

The above variable $\fb$ is not a scalar under general coordinate transformations, and cannot be  directly inserted into the potential in order to construct a generally covariant action. It has been shown, however, that a covariant formulation of the potential, of the action, and of the equations of motion  can be made compatible with the duality properties of  $(d+1)$-dimensional, homogeneous, string-frame backgrounds if  $V$ depends on the dilaton through the following non-local scalar function of $\varphi$ \cite{6,7}:
\beq
\xi(x)\equiv \xi[\varphi(x)]= -\ln \int {\d^{d+1}y\over \ls^d} \sqrt{-g(y)} ~\e^{-\varphi(y)} \sqrt{\left|\pa_\mu\varphi(y) \pa^\mu \varphi(y)\right|}~ \da \left[\varphi(x)-\varphi (y)\right].
\label{11}
\eeq
The string-length factor, $\ls^d$, is introduced here to make $\xi$ dimensionless. The variable $\xi$, unlike $\fb$, transforms as a scalar under general coordinate transformations, and exactly reduces to $\fb$ in the case of homogeneous and spatially flat metric backgrounds with  spatial sections of finite comoving volume ($\int \d^d y =V_d=$ const $<\infty$).
It has been shown, in particular, that an effective potential $V=V(\xi)$ can be induced by dilatonic loop corrections in (higher-dimensional) manifolds with compact spatial sections \cite{7}. The same non-local variable $\xi$ preserves the duality symmetry of the cosmological equations even if the dilaton is directly and non-minimally coupled to the matter sources, as recently pointed out in \cite{8}. 

In this paper we will assume that the dilaton is non-locally coupled to the gravitational sources appearing in the matter Lagrangian (or, at least, to a significant fraction of these sources), and that such a non-local interaction can be parametrized by the same variable $\xi$ used to construct a generally covariant and duality-invariant potential. We will use this approach -- motivated by the fundamental string theory symmetries -- to study the possible effects of non-locality in a simple model of ``coupled quintessence"  based on the interaction of the dilaton with the cosmic dark-matter fluid \cite{9}. The resulting effective model of ``non-local dark energy" is -- to the best of our knowledge -- qualitatively different from all models of coupled dark energy proposed so far in the literature \cite{10,10a,11,11a,11b}, and suggests new (and potentially interesting) scenarios for the discussion of the ``coincidence problem" \cite{12} and the future evolution of the currently accelerated cosmic regime. 

The paper is organized as follows. In Sect. \ref{sec2} we derive the full  set of  integro-differential, general-covariant equations of motion, in the Einstein frame, for a canonically normalized scalar field non-minimally and non-locally coupled to  the matter sources through the duality-motivated variable $\xi$ of Eq. (\ref{11}). The homogeneous limit of such equations, with perfect fluid sources, is applied in Sect. \ref{sec3} to a model of coupled dark energy based on the dilaton, and characterized by an asymptotic regime with frozen ratio of the dark-energy to the dark-matter density \cite{9,13}. It is shown, in particular, that because of non-local effects the dark-matter fluid may develop an effective pressure which tends to compensate, asymptotically, the accelerating action of the dilaton potential, eventually driving the Universe back to the standard (dust-dominated) decelerated regime. This effect is also graphically illustrated in Sect. \ref{sec4} using a simple numerical example. The main differences between uncoupled, locally coupled, and non-locally coupled models of dark energy are briefly summarized in Sect. \ref{sec5}. 

Finally, our conventions will be as follows: diag $g_{\mu\nu}= (+,-,-,-, \cdots)$, $R_{\mu\nu\a}\,^\b= \pa_\mu \Ga_{\nu \a}\,^\b+ \Ga_{\mu \r}\,^\b \Ga_{\nu\a}\,^\r- (\mu \leftrightarrow \nu)$, and $R_{\nu\a}=R_{\mu\nu\a}\,^\mu$. 

\section {Non-local matter-dilaton interactions}
\label{sec2}

Since in this paper we are not concerned with the duality properties of the string effective action, we will phenomenologically introduce the non-local coupling working directly in the Einstein frame. Also, we will use here a very simple model of matter-dilaton interactions in which the non-local terms only appear in the matter Lagrangian ${\cal   L}_m$: no additional contribution due to non-locality is present, neither in the potential nor in the possible dilaton-loop corrections to the kinetic part of the  action (see \cite{8} for the discussion of more general coupling schemes). 

We shall thus consider the (Einstein-frame) action for a canonically normalized scalar field $\phi$ (conventionally called ``dilaton"), minimally coupled to gravity and non-locally interacting with the matter  sources as follows, 
\beq
S= \int \d^{d+1}~ \sqrt{-g} \left[-{R\over 2 \lp^{d-1}}+{1\over 2} \nabla_\mu \phi \na^\mu \phi -V(\phi)\right] 
+ \int \d^{d+1}x~ \sqrt{-g}~ {\cal L}_m(\e^{-\xi}),
\label{21}
\eeq
where $\xi$ is a scalar, dimensionless, non-local function of $\phi$  defined by:
\beq
\e^{-\xi(x)}= \lp^{-d} \int \d^{d+1}y \left(\sqrt{-g}\, \e^{-\phi/\mu} \sqrt{\ep (\na \phi)^2}\right)_y \da(\phi_x-\phi_y).
\label{22}
\eeq
Here $\lp= (8\pi G)^{1/(d-1)}$ is the Planck length, and $\mu$ is a dimensional factor with the same canonical dimensions as the scalar field $\phi$. If, in particular, $\mu= [(d-1)\lp^{d-1}]^{-1/2}$, then the dimensionless ratio $\phi/\mu$ exactly corresponds to the string-theory dilaton $\varphi$, rescaled from the string to the Einstein frame (see e.g. \cite{7}). In this paper we will adopt however a phenomenological approach, leaving open the value of $\mu$ as a free, model-dependent parameter, and discussing how the final results may depend on the specific value assigned to $\mu$. Note that in Eq.(\ref{22}) we have also explicitly inserted the factor 
\beq
\ep= {\rm sign}\{ (\na \phi)^2 \}, ~~~~~~~~~~ (\na \phi)^2 \equiv \nabla_\mu \phi \na^\mu \phi,
\label{23}
\eeq
which is required in order to include into our formalism both time-like, $(\na \phi)^2>0$, and space-like, $(\na \phi)^2<0$, dilaton gradients. Finally, we are using the convenient notation in which an index appended to round brackets, $( \dots )_x$, means that all quantities inside the brackets are functions of the appended variable (similarly, $\phi_x \equiv \phi(x)$). 

The general-covariant equations of motion for this non-local model of scalar-tensor gravity can now be deduced by computing the functional derivatives of the action with respect to $g_{\mu\nu}$ and $\phi$. For the metric we obtain, separating local and non-local contributions, 
\beq
{\da S \over \da g^{\mu\nu}(x)}= \sqrt{-g} \left[-{1\over 2 \lp^{d-1}} G_{\mu\nu}+ {1\over 2} T_{\mu\nu} +{1\over 2} \left(\na_\mu \phi \na_\nu \phi -{1\over 2} g_{\mu\nu} \na \phi^2\right) +{1\over 2} g_{\mu\nu} V\right] +M_{\mu\nu}(x).
\label{24}
\eeq
Here $G_{\mu\nu}$ is the Einstein tensor, $T_{\mu\nu}$ represents the (local) stress-tensor contribution of the matter action, defined as usual by
\beq
{1\over 2} \rg \,T_{\mu\nu} = {\da (\rg \cl) \over \da g^{\mu\nu}},
\label{25}
\eeq
and $M_{\mu\nu}$ represents a new, non-local contribution due to the explicit presence of the metric in the definition of $\xi$, according to Eq. (\ref{22}). In particular,
\beq
M_{\mu\nu}(x)= \int \d^{d+1} y \left( \rg \cl'\right)_{y} 
{\da\over \da g^{\mu\nu}(x)} \e^{-\xi(y)},
\label{26}
\eeq
where $\cl'$ is the derivative of the matter Lagrangian with respect to its argument $\exp(-\xi)$,
\beq
\cl'\equiv {\pa \cl \over \pa(\e^{-\xi})}=-\e^\xi\, {\pa \cl\over \pa \xi},
\label{27}
\eeq
and (from Eq. (\ref{22}))
\beq
{\da\over \da g^{\mu\nu}(x)} \e^{-\xi(y)}=-{1\over 2} \lp^{-d} \left(\ga_{\mu\nu} \rg \e^{-\phi/\mu} \rp \right)_x\da(\phi_{y}-\phi_x),
\label{28}
\eeq
where
\beq
\ga_{\mu\nu} = g_{\mu\nu} - {\na_\mu \phi \na_\nu \phi \over (\na \phi)^2}.
\label{29}
\eeq
Inserting these results into Eq. (\ref{26}), summing up all terms of Eq. (\ref{24}), multiplying by $2 \lp^{d-1} / \rg$, and imposing the condition of stationary action, $\da S/\da g^{\mu\nu}=0$, we are led  to the generalized (integro-differential) gravitational equations:
\beq
G_{\mu\nu} = \lp^{d-1}\left[T_{\mu\nu} +\na_\mu \phi \na_\nu \phi -{1\over 2} g_{\mu\nu} (\na \phi)^2 + g_{\mu\nu} V -\ga_{\mu\nu}  \e^{-\phi/\mu} \rp I_m\right],
\label{210}
\eeq
where
\beq
I_m= \lp^{-d} \int \d^{d+1} y \left( \rg \cl'\right)_{y} \da(\phi_{y}-\phi_x).
\label{211}
\eeq

Similarly, the functional derivative of the action (\ref{21}) with respect to the dilaton $\phi$ contains both local and non-local contributions,
\beq
{\da S\over \da \phi(x)}= - \rg \left( {\pa V\over \pa \phi} + \na^2 \phi \right)_x + B(x),
\label{212}
\eeq
where $\na^2 \phi= \na_\mu\na^\mu \phi$, and where the non-local contribution $B(x)$ is due to the explicit dependence of the variable $\xi$ on the dilaton:
\beq
B(x)= \int \d^{d+1} y \left( \rg \cl'\right)_{y} 
{\da\over \da \phi(x)} \e^{-\xi(y)}. 
\label{213}
\eeq
A direct computation of $B(x)$ (see Appendix A), inserted into the condition $\da S/ \da \phi=0$,  finally leads to the generalized (integro-differential) scalar-field equation of motion,
\beq
\na^2 \phi+{\pa V \over \pa \phi} + \ep {\e^{-\phi/\mu}\over \rp} \, \ga_{\mu\nu} \na^\mu \na ^\nu \phi \, I_m+
\cl'\left({\e^{-\xi}\over \mu} - \e^{-\phi/\mu} J\right)=0,
\label{214}
\eeq
where 
\beq
J(x)= \lp^{-d} \int \d^{d+1} y \left( \rg \,\rp\right)_y \da'(\phi_x-\phi_y),
\label{215}
\eeq
and where $\da'$ denotes the derivative of the delta function with respect to its argument. 

For the cosmological applications of this paper we need to specify the generally covariant, non-local set of coupled equations (\ref{210}), (\ref{214}) to the simple case of a homogeneous, isotropic, spatially flat, $d=3$ background, sourced by a perfect-fluid stress tensor. In the cosmic time gauge: 
\bea
&&
g_{\mu\nu}= {\rm diag}(1, - a^2 \da_{ij}), ~~~~~~a= a(t), ~~~~~~ \phi=\phi(t),\nonumber \\
&&
T_\mu\,^\nu=  {\rm diag}(\r, - p \da_i^j), ~~~~~~~ \r=\r(t), ~~~~~~ p=p(t).
\label{216}
\eea
In this simple case $ \ep=1$, $\ga_{00}=0$, $\ga_i^j= \da_i^j$, and we find, as a first result, that the $(00)$ component of Eq. (\ref{210}) reduces to the standard Friedman equation,
\beq
6 H^2= 2 \lp^2 \left(\r+{\fpu \over 2} +V\right).
\label{217}
\eeq

Concerning the spatial components we note, first of all, that in the homogeneous limit the non-local variable (\ref{22}) reduces to
\beq
\e^{-\xi}~ \ra~ \Om_3\, a^3 \e^{-\phi/\mu} \equiv \e^{\fb},
\label{218}
\eeq
where $\Om_3= \lp^{-3} \int \d^3x$ is  the constant volume factor (in Planck units) of the compact spatial sections of the manifold (\ref{216}). The non-local term of Eq. (\ref{210}) becomes, in the same  limit,
\beq
 \e^{-\phi/\mu} \rp\, I_m ~ \ra ~ \e^{-\fb} \cl' = - {\pa \cl \over \pa \fb}.
 \label{219}
 \eeq
Thus, from the $(ij)$ components of Eq. (\ref{210}) we obtain the spatial equation
 \beq
 4 \dot H + 6 H^2= 2 \lp^2 \left[ -p-\left( {\fpu^2 \over 2} -V\right) + 
{\pa \cl \over \pa \fb} \right],
\label{220}
\eeq
where the last term on the right-hand side represents the only correction induced by non-locality with respect to the standard Einstein equation. 

We shall now consider the dilaton equation (\ref{214}). In the homogeneous  background (\ref{216}) there is a big simplification, due to the fact that
\bea
&&
\ep {\e^{-\phi/\mu}\over \rp} \, \ga_{\mu\nu} \na^\mu \na ^\nu \phi \, I_m 
 ~~ \ra ~ -{3H \over \fpu}\, {\pa \cl \over \pa \fb},
\nonumber \\ &&
 -\cl' \e^{- \phi/\mu} J ~~ \ra ~ {3H \over \fpu}\, {\pa \cl \over \pa \fb}.
\label{221}
\eea
These two non-local terms thus exactly cancel each other in the homogeneous limit of Eq. (\ref{214}), and we  are left with the dilaton equation
\beq
\fpp+ 3 H \fpu +{\pa V \over \pa \phi} -{1\over \mu}  {\pa \cl \over \pa \fb}=0,
\label{222}
\eeq
where the last term represents the correction induced by the explicit  dependence of $ \cl$ on $\fb$. 

In view of our subsequent discussion it is finally convenient to rewrite the full set of cosmological equation using an effective fluido-dynamical representation of the scalar field source. Let us introduce, to this purpose, the effective energy density and pressure $\r_\phi$, $p_\phi$, and the effective ``scalar charge" density of the matter sources, $\sg$, defined by 
\beq
\r_\phi= {\fpu^2\over 2} + V, ~~~~~~~~~~
p_\phi= {\fpu^2\over 2} - V, ~~~~~~~~~~
{\sg \over 2} =-  {\pa \cl \over \pa \fb}. 
\label{223}
\eeq
Our cosmological equations  (\ref{217}), (\ref{220}), (\ref{222}) 
can then be rewritten respectively as:
\bea
&&
6H^2= 2 \lp^2 \left(\r+\r_\phi\right),
\label{224} \\ &&
4 \dot H+ 6 H^2= - 2 \lp^2\left(p+p_\phi+{\sg\over 2} \right), 
\label{225} \\ && 
\dot \r_\phi+ 3H\left(\r_\phi+p_\phi \right) = - {\sg \over 2 \mu} \fpu,
\label{226}
\eea
and their combination gives the covariant conservation equation of the matter energy density:
\beq
\dot \r + 3H\left( \r+ p+ {\sg\over 2}\right) = {\sg \over 2 \mu} \fpu.
\label{227}
\eeq

It should be stressed,  before concluding  this section,  that the genuinely new effect of the non-local interactions is the appearance of the contribution proportional to $\sg$ on the right-hand side of the spatial equation (\ref{225}): this contribution acts as an ``additional" pressure of the matter fluid, being equivalent to the term one could obtain from the standard Einstein equations through the shift
\beq
p ~ \ra ~ p + {\sg\over 2}
\label{228}
\eeq
(see also Eq. (\ref{227})). It is true that the $\xi$-dependence of the  matter Lagrangian also induces further corrections to the Einstein equations, in particular introduces the direct (non-minimal) coupling between the dilaton and the matter sources represented by the terms $\pm \sg \fpu/2 \mu$ in Eqs. (\ref{226}), (\ref{227}). However, a similar type of coupling (triggered by the modification of the dilaton equation of motion (\ref{222})) would also be obtained in the more conventional framework  in which the dilaton is {\em locally} coupled to matter, i.e. $\cl= \cl (\phi)$: hence, those terms do not represent an important qualitative difference from previous models of coupled quintessence (see e.g. \cite{10,10a,11}). The effective pressure (\ref{228}), on the contrary, can only be obtained if the matter-dilaton interaction is directly affected by the global background geometry as in the case of the non-local coupling defined by Eq. (\ref{22}). A possible phenomenological effect of such a non-locally induced pressure will be discussed in the following section.

\section {Asymptotic freezing and pressure backreaction}
\label{sec3}

The generalized cosmological equations derived in the previous section will now be applied to a model of dark energy formulated in a string cosmology context \cite{9}, based on the dilaton and on the principle of asymptotic saturation of the loop corrections in the large ``bare coupling" limit $ \phi \ra + \infty$ \cite{14}. In such a context the post-inflationary Universe naturally evolves towards a ``late-time" regime where the dark-energy field $\phi$ keeps rolling down along an exponentially suppressed potential, and may become permanently coupled to (a significant fraction of) the dark-matter sources. 

Such a late-time scenario can be appropriately described by the generalized equations (\ref{224})--(\ref{227}) by identifying the fluid source with the fraction of coupled dark-matter component (i.e. by setting $\r= \r_m$, $p=p_m=0$), and assuming for $V$ and $\sg$ the following asymptotic parametrization:
\beq
V=V_0 \,\e^{-\phi/\mu}, ~~~~~~~~~~~
\sg= q_0\, \r_m.
\label{31}
\eeq
The constant values of $q_0$, $V_0$ are typical parameters of the asymptotic saturation regime, and are in principle calculable at the fundamental string-theory/M-theory level \cite{9,14} (in particular, the dimensionless ratio $q_0=\sg/\r_m$ represents the effective dilaton charge per unit of gravitational mass of the dark matter particles, see e.g. \cite{15}). For the discussion of this asymptotic regime  it will be convenient to separate the kinetic and potential part of the dilaton energy density, by setting
\beq
\r_\phi=\r_k+\r_V, ~~~~~~~~~~
\r_k={\fpu^2\over 2}, ~~~~~~~~~~
\r_V=V.
\label{32}
\eeq
Introducing critical units,
\beq
\Om_k= {\lp^2 \r_k\over 3 H^2},  ~~~~~~~~~~
\Om_V= {\lp^2 \r_V\over 3 H^2},  ~~~~~~~~~~
\Om_m= {\lp^2 \r_m\over 3 H^2}, 
\label{33}
\eeq
and using Eq. (\ref{31}) we can then rewrite the cosmological equations (\ref{224})--(\ref{227}) 
respectively as
\bea
&&
1=\Om_m+\Om_k+\Om_V,
\label{34} \\ &&
1+{2\dot H\over 3H^2}= \Om_V-\Om_k -{q_0\over 2} \Om_m,
\label{35}  \\ &&
\dot \r_\phi+ 6H\r_k=-{q_0\over 2 \mu} \r_m \fpu,
\label{36}  \\ &&
\dot \r_m+ 3H\r_m\left(1+{q_0\over 2}\right)= {q_0\over 2 \mu} \r_m \fpu.
\label{37}
\eea

In the case of local dilaton coupling -- i.e. in the absence of the ``induced pressure" $\sg/2$ generating the last term on the right-hand side of Eq. (\ref{35}) and the last term on the left-hand side of Eq. (\ref{37}) -- it is well known that the above set of asymptotic equations can be satisfied by a ``freezing" configuration in which $\r_m$, $\r_k$, $\r_v$, $H^2$ scale in time in the same way, so that the ratio $\r_m/\r_\phi$ turns out to be fixed \cite{9,10,10a,11}. In that limit the critical fractions $\Om_m$, $\Om_k$, $\Om_V$ are also separately fixed at constant values (determined by $\mu$ and $q_0$ only), and  a simple analysis gives
\bea
&&
\Om_k^{\rm loc}= {6 \lp^2 \mu^2\over (2+q_0)^2}, ~~~~~~~~~~~~
\Om_V^{\rm loc}={6 \lp^2 \mu^2+q_0(2+q_0)\over (2+q_0)^2},
\nonumber \\ &&
\Om_m^{\rm loc}=1-\Om_k^{\rm loc}-\Om_V^{\rm loc}
\label{38}
\eea
(the superscript ``loc" is to remind that we are solving a model of  locally-coupled dark energy). According to this solution the Universe thus becomes dominated by a mixture of dark-matter and dilaton (kinetic plus potential) energy density. The rate of expansion of this asymptotic regime is controlled by the kinematic parameter
\beq
\left(\ddot a \over a H^2\right)_{\rm loc}= 1 + 
\left(\dot H\over H^2\right)_{\rm loc}=
-{1\over 2}+{3\over 2} \left(\Om_V^{\rm loc}-\Om_k^{\rm loc}\right)
={q_0-1\over q_0+2},
\label{39}
\eeq
and the corresponding effective equation of state (defined as $w_{\rm eff}\equiv -1-2\dot H/3H^2$) is given by 
$w_{\rm eff}=-q_0/(2+q_0)$. The evolution is therefore accelerated for $q_0>1$ (the  possibility $q_0<-2$ is to be excluded, see e.g. \cite{8}). 

It can be shown, in particular, that for strong enough coupling (i.e. small enough values of $\mu$) the above configuration represents a stable attractor of the cosmological dynamics \cite{10}. 
Hence, for appropriate values of $\mu$ and of the scalar charge $q_0$, the dilaton can play the role of the dark energy field forcing the Universe to an asymptotic regime of cosmic accelerated evolution and frozen dark-matter over dark-energy density ratio \cite{9,13}. If, on the contrary, the coupling is small (namely, the value of $\mu$ is large enough) then the Universe is attracted towards a ``more standard" dark-energy dominated, uncoupled regime where $\Om_m$ can be neglected  \cite{10, 10a}, and the cosmic expansion is fully driven by $\r_\phi$, with an effective equation of state $w_{\rm eff}=-1+(3 \lp^2 \mu^2)^{-1}$. This different type of asymptotic configuration is thus accelerated for $\lp |\mu|>1/\sqrt 2$.

In order to discuss the effects of the non-local coupling we will now concentrate on the asymptotic freezing configuration described by the system (\ref{34})--(\ref{37}) with all terms (including $\Om_m$ and the induced pressure $\sg/2$) taken into account. Let us look for freezing solutions by requiring for  $\r_m$, $\r_k$, $\r_v$ the same scaling behavior and thus imposing, as a first condition, that 
\beq
{\dot \r_m\over \r_m}= {\dot \r_\phi\over \r_\phi}.
\label{310}
\eeq
Using Eqs. (\ref{34}), (\ref{36}), (\ref{37}) we obtain:
\beq
{\fpu \over H}= {6 \mu \over q_0} \left[\Om_V\left(1+{q_0\over 2}\right)
-\Om_k\left(1-{q_0\over 2}\right)\right].
\label{311}
\eeq
We also impose, as a second condition, that 
\beq
{\dot \r_m\over \r_m}= {\dot \r_V\over \r_V}.
\label{312}
\eeq
Using Eqs. (\ref{31}), (\ref{32}), (\ref{37}) we obtain:
\beq
{\fpu \over H}=3\mu.
\label{313}
\eeq
The combination of Eqs. (\ref{311}), (\ref{313}), and the use of the definition $\Om_k= \lp^2 \fpu^2/6H^2$, finally gives the result
\bea
&&
\Om_k={3\over2} \lp^2 \mu^2, ~~~~~~~~~~~~~~~~~
\Om_V={2 q_0+ 3(2-q_0)\lp^2 \mu^2\over 2 (2+q_0)},
\nonumber \\ &&
\Om_m=1-\Om_k-\Om_V.
\label{314}
\eea
Thus, even in case of non-local coupling our  equations admit solutions describing a freezing configuration, dominated by a mixture of  dark-matter, dilaton-kinetic, and dilaton-potential energy density, and characterized by $\r_m/\r_\phi=$ const (just as in the case of local coupling). 

There is, however, a surprising result if we compute the acceleration parameter of this asymptotic regime. Using Eq. (\ref{35}), and the results (\ref{314}) for the critical fractions of energy density, we obtain
\beq
{\ddot a \over a H^2}= 1 + 
{\dot H\over H^2}=
-{1\over 2}+{3\over 2} \left(\Om_V-\Om_k-{q_0\over 2} \Om_m\right)
\equiv -{1\over 2},
\label{315}
\eeq
i.e. a constant negative result quite independent of the values of $q_0$ and $\mu$ ! (compare with Eq. (\ref{39})). The integration of the above equation for $\dot H$ gives, in particular, 
\beq
a \sim t^{2/3}, ~~~~~~~~~~~~
H^2 \sim \r_m \sim \r_\phi \sim a^{-3}.
\label{316}
\eeq
This asymptotic freeezing phase has thus the same decelerated behavior as the standard cosmological phase describing dust-matter dominated evolution. 

This result can be understood by noting that Eqs. (\ref{315}) and (\ref{35}) imply
\beq
-w_{\rm eff}\equiv
\Om_V-\Om_k-{q_0\over 2} \Om_m= -{\lp^2\over 3H^2} \left (p_\phi+ {\sg\over 2}\right)\equiv0,
\label{317}
\eeq
namely a zero total pressure for the dilaton-dark matter system. This means, in other words, that the effective dark-matter pressure $\sg/2$ (induced by the non-local coupling) generates asymptotically a backreaction which exactly compensates -- at least in this context -- the dilaton pressure $p_\phi$, leading the system to restore the standard dust-matter configuration. This effect will be further illustrated in the next section through a simple numerical integration of the coupled cosmological equations. 

It should be noted, finally, that the above asymptotic configuration must  satisfy various conditions in order to be realistic (for instance, $0 \leq \Om_k\leq1$, $0\leq \Om_V \leq 1$), and to represent a stable attractor of the cosmological dynamics. In our case, it can be shown \cite{18a} that all those conditions imply $\lp |\mu| <1/\sqrt 3$. With larger values of $\mu$ (i.e. smaller values of the dilaton-dark matter coupling) the Universe is attracted towards the same final configuration one would obtain in the case of local coupling: the effective equation of state becomes $w_{\rm eff}=1+(3 \lp^2 \mu)^{-1}$ and, as before, this final regime may be accelerated provided $\lp 
|\mu|>1/\sqrt 2$. In that case, indeed, the dark-matter contribution to the cosmological dynamics becomes sub-leading, so that also the backreaction of the non-local pressure becomes ineffective.

\section {A numerical example}
\label{sec4}

For an explicit illustration of the smooth background evolution towards the final freezing regimes (\ref{38}) and (\ref{314})  we will adopt here the same model of time-dependent dilaton charge already used in previous papers \cite{9,13}, parametrized by
\beq
\sg= q(\phi) \, \r_m, ~~~~~~~~~~~~~
q(\phi)= q_0{\e^{q_0\phi}\over c^2 + \e^{q_0\phi}}.
\label{41}
\eeq
This form of $\sg$ is motivated by the principle of asymptotic saturation of the dilaton loop corrections (note that $q \ra q_0$ for $\phi \ra \infty$), which provides for the dimensionless parameter $c^2$ a typical value of order $10^2$ -- i.e. of the order of the number of gauge bosons  contained into the fundamental GUT group \cite{14}. 
In addition, we  will complete our model by adding to the cosmic fluid the required fraction of dust baryonic matter $\r_b$, $p_b=0$, uncoupled to the dilaton, and thus satisfying the standard energy conservation equation. By setting $\r=\r_m+\r_b$, $p=0$, our basic set of equations (\ref{224})--(\ref{227}) thus reduces to
\bea
&&
6H^2= 2 \lp^2 \left(\r_m+\r_b+\r_\phi\right),
\label{42} \\ &&
4 \dot H+ 6 H^2= - 2 \lp^2\left(p_\phi+{q\over 2}\r_m \right), 
\label{43} \\ && 
\dot \r_\phi+ 3H\left(\r_\phi+p_\phi \right) = - {q \over 2 \mu}\r_m \fpu,
\label{44}\\ &&
\dot \r_m + 3H\r_m\left( 1+  {q\over 2}\right) = {q \over 2 \mu}\r_m \fpu, 
\label{45}\\ &&
\dot \r_b + 3H\r_b=0, 
\label{46}
\eea
where  $q=q(\phi)$ is given by Eq. (\ref{41}), and the dilaton potential is given by Eq. (\ref{31}). We will assume, for a realistic scenario, that the critical fraction of baryon energy density is small, and thus represents  a negligible perturbation of the asymptotic configuration (\ref{314}). 

\begin{figure}[t]
\begin{center}
\includegraphics[width=100mm]{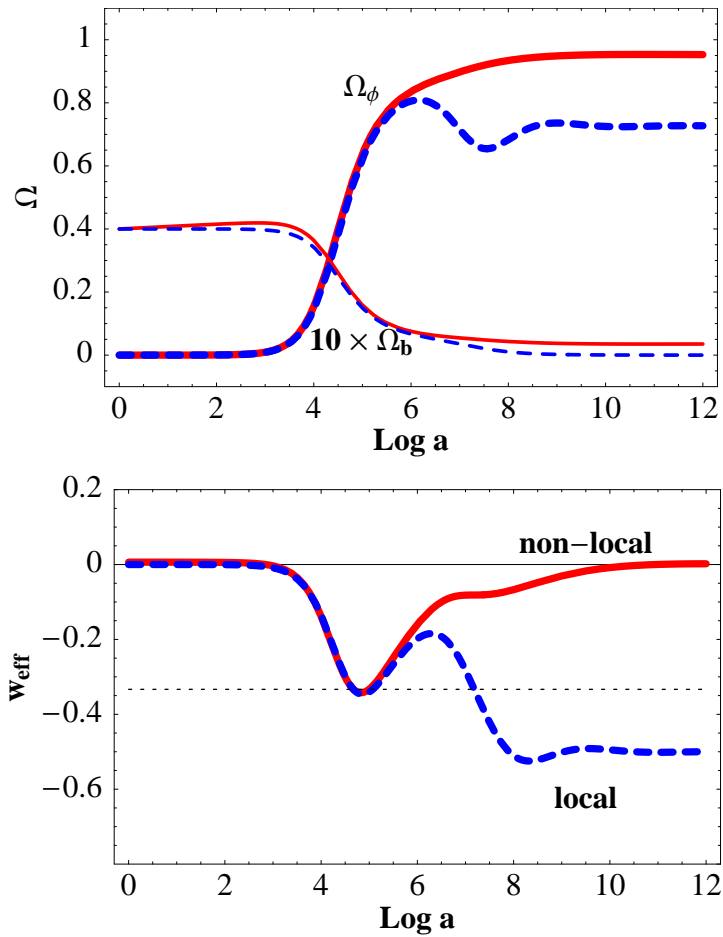}  
\end{center}
\caption{\sl Top panel: late-time scaling evolution of the critical fraction of baryon energy density multiplied by ten, and of the critical fraction of dilaton dark-energy, for the case of local dilaton coupling (dashed curves)  and non-local coupling (solid curves). The plots are obtained through a numerical integration of the set of equations (\ref{42})--(\ref{46}) with $q=q(\phi)$  given by Eq. (\ref{41}) and the dilaton potential  given by Eq. (\ref{31}). 
Bottom panel: evolution of the total effective equation of state (\ref{47}) for the case of local coupling (dashed  curve) and non-local coupling (solid curve). The horizontal dotted line corresponds to $w_{\rm eff}=-1/3$, and marks the border of the accelerated regime.} 
\label{fig1}
\end{figure}

The above system of coupled differential equations has been numerically integrated with appropriate initial conditions, using Planck units $2 \lp^2=1$, and using the values $q_0=2$, $c^2=150$, 
$\mu= 0.55$, $V_0=3\times 10^{-6}$.  Note that the chosen value of $\mu$ satisfies the stability condition of the freezing regime (see the discussion at the end of Sect. \ref{sec3}). 
The results of the numerical integration are illustrated in Fig. \ref{fig1}, where we have reported  the scaling evolution $\Om=\Om(a)$ of various cosmic components, and the total effective equation of state, for both the local and the non-local model.

In the top panel of Fig. \ref{fig1}  the thin solid  curve represents the critical fraction of baryon energy density $\Om_b= \lp^2 \r_b/3H^2$, multiplied by ten for a better graphical illustration, in the case of non-local dilaton coupling. The bold solid curve represents the evolution of  $\Om_\phi=\Om_k+\Om_V$ for the case of non-local coupling (i.e.  including the induced pressure $q\r_m/2$ in the cosmological equations). Finally, the dashed curves represent the same quantities
for the case of local coupling (i.e. neglecting the induced pressure). 
Note that baryons are  uncoupled to the dilaton, and thus obey the standard scaling behavior $\r_b \sim a^{-3}$; however, the behavior of $\Om_b$ is different in the cases of local and non-local coupling because of the different evolution of $H^2$. 
Note also that the beginning of the asymptotic freezing regime  is triggered by the rise of $q(\phi)$ (not shown in the picture) and of the dilaton potential energy, which starts to become dominant at some epoch determined by the potential amplitude $V_0$.

We recall that, as discussed in Sect. \ref{sec3}, the local dilaton-dark matter coupling leads the Universe to an asymptotic regime where the ratio $\r_m/\r_\phi$ is frozen and, for appropriate values of $q_0$ (see Eq. (\ref{39})), the kinematics describes a phase of eternal accelerated expansion. If we include non-local effects (according to the model considered in this paper) the freezing regime is still asymptotically established; however, the associated kinematics is exactly the same as that of the dust-dominated, decelerated regime. A limited period of accelerated expansion (possibly reproducing the currently observed large-scale evolution) can exist. However, such acceleration is doomed to disappear with the rise of the non-local pressure, which eventually compensates the accelerating action of the dilaton potential driving to zero the total effective cosmic pressure. 

For a better illustration of this last point we have also plotted, in the bottom panel of Fig. \ref{fig1}, the effective equation of state obtained from Eqs. (\ref{42}), (\ref{43}),
\beq
w_{\rm eff}=-1-{2 \dot H\over 3 H^2}= {p_\phi+{q}\r_m/2 \over \r_m+ \r_b+ \r_\phi}.
\label{47}
\eeq
As before, the dotted curve corresponds to the local coupling (without the term $q\r_m/2 $), and the solid curve corresponds to the non-local coupling (all terms included). In both cases the cosmological system evolves from the initial condition $w_{\rm eff}=0$, associated to a dark-matter dominated state in which the dilaton is still sub-leading.

When the dilaton comes into play the total effective pressure tends to become negative, leading the Universe towards the accelerated regime where $w_{\rm eff} <-1/3$. In the case of local coupling the Universe asymptotically reaches an asymptotic freezing regime with $w_{\rm eff}= -q_0/(2+q_0)$. In the case of non-local coupling, on the contrary, the Universe crosses the acceleration boundary for a very short period, and then bounces back to an asymptotic state of dust-dominated evolution, with  $w_{\rm eff}=0$. The duration and the position of the accelerated phase depends on the parameters of the coupled dark-energy scenario.

\section {Conclusion}
\label{sec5}

It seems appropriate to conclude this paper by comparing the possible impact of different dark-energy scenarios (in particular, the uncoupled, the locally coupled, and the non-locally coupled scenario) upon the so-called problem of the cosmic coincidence \cite{12}. 

In models where the dark-energy field is asymptotically uncoupled to dark matter (see e.g. \cite{16}) the final regime is accelerated, and dominated by the potential energy $\r_\phi$ of a scalar (``quintessence") field: the dark-matter density $\r_m$ is diluted faster than $\r_\phi$, and is doomed to become subdominant. Thus, the current approximate equality $\r_m \sim \r_\phi$ turns out to be a coincidence of the present epoch. The acceleration, however, is not a coincidence, in the sense that it is typical of an epoch of infinite time extension (once started, it last forever). 

In models where dark energy and dark matter are locally (and strongly enough) coupled (see e.g. \cite{9,10,10a,11}) the final asymptotic regime is accelerated, and characterized by a frozen value of $\r_m/\r_\phi$. Thus, the relation $\r_m \sim \r_\phi$ and the positive value of the acceleration are both implemented during an epoch of inifinite time extension and, in this sense, they are not a coincidence of the present epoch. 

In models where dark energy and dark matter are non-locally (and strongly enough) coupled, as in the example of this paper, the final asymptotic regime is characterized by a frozen value of $\r_m/\r_\phi$, but it is decelerated. Thus, the relation $\r_m \sim \r_\phi$ is not a coincidence of the present epoch; the acceleration, however, is localized within an epoch of small finite time extension, so that the cosmic coincidence persists (possibly in relaxed form, as the accelerated epoch is not a point in phase space). It may be noted that such a situation is exactly complementary to the one of the uncoupled scenario. 

It should be stressed, finally, that the three above scenarios are characterized by a different cosmological dynamics not only during the final asymptotic regime, but also during the epoch preceding the beginning of the accelerated regime. We may hope, therefore, that direct/indirect astrophysical observations, providing us with more and more accurate information about the past cosmological history, will also tell us which of the above scenarios can eventually provide a better description of our Universe.

\bigskip
{\it Acknowledgment.} We are grateful to Gabriele Veneziano for helpful discussions during the early stages of this work. 

\begin{appendix}
\renewcommand{\theequation}{A.\arabic{equation}}
\setcounter{equation}{0}
\section{}

In this Appendix we present an explicit computation of the functional derivative of the (non-local) matter action with respect to the dilaton, and in particular of the scalar functional $B(x)$ defined by Eq.  (\ref{213}). Using the definition (\ref{22}) of $\xi(x)$ we obtain
\bea
{\da\over \da \phi(x)} \e^{-\xi(x')}&=&\lp^{-d} \int \d^{d+1}y \Bigg\{-{1\over \mu} \left( \rg \e^{-\phi/\mu} \rp \right)_y \da (\phi_{x'}-\phi_y) \da^{d+1}(x-y)
\nonumber \\ &+&
 \left(\rg \e^{-\phi/\mu} \rp \right)_y \da' (\phi_{x'}-\phi_y) 
\left[ \da^{d+1}(x-x') - \da^{d+1}(x-y)\right]
\nonumber \\ &-&
\pa_\mu \left[{\rg \e^{-\phi/\mu} \ep\, \pa^\mu \phi \over \rp} \da (\phi_{x'}-\phi_y)\right]_y \da^{d+1}(x-y)
 \Bigg\},
\label{A1}
\eea
where $\pa_\mu = \pa/\pa y^\mu$, and $\da'$ denotes the derivative of the delta function with respect to its argument. The first term of this integral exactly cancels the term containing $\pa_\mu [\exp(-\phi/\mu)]$ in the last part of the integral; also, the third term exactly cancels with the term containing $\pa_\mu [\da(\phi_{x'}-\phi_y)]$ in the last part of the integral. Thus, we are left with:
\bea
{\da\over \da \phi(x)} \e^{-\xi(x')}&=&\lp^{-d}~ \da^{d+1} (x-x') \int \d^{d+1}y \left(\rg \e^{-\phi/\mu} \rp \right)_y \da' (\phi_{x'}-\phi_y) 
\nonumber \\ &-& 
\ep \lp^{-d} \e^{-\phi/\mu} \da(\phi_{x'}-\phi_x)\pa_\mu  \left( \rg \pa^\mu \phi\over \rp\right)_x.
\label{A2}
\eea
The second term on the right-hand side of the above equation can now be conveniently rewritten as 
\bea
&&
-\ep \lp^{-d} \e^{-\phi/\mu} \da(\phi_{x'}-\phi_x)\rg ~ \nabla_\mu  \left( \pa^\mu \phi\over \rp\right)_x=
\nonumber \\ &&
=- \ep \lp^{-d} \e^{-\phi/\mu} \da(\phi_{x'}-\phi_x){\rg \over \rp} \ga_{\mu\nu} \na^\mu \na^\nu \phi,
\label{A3}
\eea
where $\ga_{\mu\nu}$ is defined in Eq. (\ref{29}). The first term on the right-hand of Eq. (\ref{A2}), containg $\da'$, can be rewritten using the properties of the delta distribution, and exploiting the identities:
\beq
\d y_0= {\d \phi_y\over \dot \phi_y}, ~~~~~~~~~~
{\d \over \d \phi_y}= {1\over \dot \phi_y} {\d \over \d y_0}.
\label{A4}
\eeq
We obtain:
\bea
&&
\lp^{-d}~ \da^{d+1} (x-x') \int \d^{d+1}y \left(\rg \e^{-\phi/\mu} \rp \right)_y \da' (\phi_{x'}-\phi_y)=
\nonumber \\ &&
~~~~~~~
=\da^{d+1}(x-x') \left[-{\e^{-\xi(x)}\over \mu} + \e^{-\phi(x)/\mu} J(x)\right],
\label{A5}
\eea
where $J$ is the integral defined in Eq. (\ref{215}). Summing the two terms of Eq. (\ref{A2}), inserting the result into Eq. (\ref{213}) (with $x'=y$), and integrating over $y$, we obtain:
\beq
B(x)= - \left(\rg \cl'\right)_x \left({\e^{-\xi}\over \mu} - \e^{-\phi/\mu} J\right)_x- \ep \left({\e^{-\phi/\mu} \rg \over \rp}
\ga_{\mu\nu} \na^\mu \na^\nu \phi ~ I_m
 \right)_x.
 \label{A6}
 \eeq
Summing this contribution to the local terms of Eq. (\ref{212}) we are finally  led to the generalized dilaton equation (\ref{214}). 

\end{appendix}


\begin{thebibliography}{99}
\frenchspacing
\newcommand{\bb}{\bibitem}

\bb{1}A. A. Tseytlin, Mod. Phys. Lett. A {\bf 6}, 
1721 (1991).

\bb{2}G. Veneziano, Phys. Lett. B {\bf 265}, 287  (1991).

\bibitem{3} K. A. Meissner and G. Veneziano, Mod. Phys. Lett. A 
{\bf 6}, 3397 (1991);  Phys. Lett. B {\bf 267},  33 (1991). 

\bibitem{4}M.  Gasperini, J. Maharana and G. Veneziano, Phys.  
Lett. B {\bf 272}, 277 (1991); Nucl. Phys. B 
{\bf 472}, 349  (1996). 

\bb{5}M. Gasperini and G. Veneziano, Phys. Lett. B {\bf 277},
256 (1992).

\bb{6}M. Gasperini, M. Giovannini and G. Veneziano, Phys.  
Lett. B {\bf 569}, 113 (2003). 

\bb{7}M. Gasperini, M. Giovannini and G. Veneziano,  Nucl. Phys. B
{\bf 694}, 206  (2004). 

\bb{8}M. Gasperini, {\em Dilaton cosmology and phenomenology}, Lect. Notes Phys. {\bf 737}, p. 789, in press (arXiv:hep-th/0702166).

\bb{9}M. Gasperini, F. Piazza and G. Veneziano,  Phys. Rev. D {\bf 65}, 023508 (2001).

\bb{10} L. Amendola, { Phys. Rev.} D {\bf 62},  043511 (2000). 

\bb{10a}L. Amendola and D. Tocchini-Valentini, Phys. Rev. D {\bf 64}, 043509 (2001);  Phys. Rev. D {\bf 66},  043528 (2002).

\bb{11}L. P. Chimento, A. S. Jakubi and D. Pavon, Phys. Rev. 
D \textbf{62}, 063508 (2000); W. Zimdahl, D. J. Schwarz, A. B. Balakin 
and D. Pavon, Phys. Rev. D {\bf 64}, 063501 (2001);  S. Sen and A. A. Sen, Phys. Rev. D \textbf{63},124006 (2001); A. A. Sen and S. Sen, 
 Mod. Phys. Lett. A {\bf 16}, 1303 (2001); W. Zimdahl and D. Pavon, Phys. Lett. B {\bf 521}, 133 (2001).

\bb{11a}J. Grande, J. Sol\`a and H. Stefancic, JCAP {\bf 0608}, 011 (2006).

\bb{11b}A Fuzfa and J. M. Alimi, Phys. Rev. D {\bf 75}, 123007 (2007). 

\bb{12}P. Steinhardt, in {\em Critical problems in physics}, ed. by V. L. Fitch and D. R. Marlow (Princeton University press, Princeton NJ, 1997).

\bb{13} L. Amendola, M. Gasperini, D. Tocchini-Valentini and C. Ungarelli,   Phys. Rev. D {\bf 67}, 043512  (2003);  L. Amendola, M. Gasperini and  F. Piazza,  JCAP {\bf 09} , 014 (2004);  
Phys. Rev. D {\bf 74}, 127302 (2006). 

\bb{14}G. Veneziano, JHEP {\bf 0206}, 051 (2002).

\bb{15}M. Gasperini, {\em Elements of string cosmology} (Cambridge University Press, Cambridge, 2007), Chapter 9. 

\bb{18a}L. Amendola et al., in preparation. 

\bb{16}B. Ratra and P. J. E. Peebles, Phys. Rev. D {\bf 37}, 3406 (1988);  C. Wetterich, Nucl. Phys. B {\bf 302}, 668 (1988); M. S. Turner and M. White,  Phys. Rev. D {\bf 56}, R4439 (1997); R. R. Caldwell, R. Dave and P. J. Steinhardt, Phys. Rev. Lett. {\bf 80}, 1582  (1998); I. Zlatev, L. Wang and P. J. Steinhardt, Phys. Rev. Lett. {\bf 82}, 896 (1999); P. J. Steinhardt, L. Wang and  I. Zlatev, 
Phys. Rev. D {\bf 59}, 123504 (1999). 
 

\end{thebibliography}
\end{document}